\documentclass[11pt]{article}
\usepackage{cite} 

\newenvironment{proof}{\noindent {\em Proof}.\ }{\hspace*{\fill}$\halmos$\medskip}

\usepackage{amsfonts,amssymb, amsmath, curves}

\usepackage{graphicx}
\usepackage{makeidx}     
\usepackage{graphicx}    

\newtheorem{theorem}{Theorem}
\newtheorem{itlemma}{Lemma} 
\newtheorem{itproposition}[itlemma]{Proposition}
\newtheorem{itcorollary}[itlemma]{Corollary}
\newtheorem{itremark}[itlemma]{Remark}
\newtheorem{itdefinition}[itlemma]{Definition}
\newtheorem{itexample}[itlemma]{Example}

\newenvironment{lemma}{\begin{itlemma}\rm}{\end{itlemma}} 

\newcommand{\halmos}{\rule{1ex}{1.4ex}}

\newcommand{\arrowschem}[2]{\raisebox{-2ex}%
	{$\stackrel{\stackrel{\displaystyle#1}{\longrightarrow}}%
	{\stackrel{\longleftarrow}{#2}}$}}

\newcommand{\s}{s}
\newcommand{\es}{c}
\newcommand{\fs}{d}
\newcommand{\e}{e}
\newcommand{\f}{f}

\date{}

\begin{document}

\title{A remark on the number of steady states in a multiple futile cycle}
\author{Liming~Wang and~Eduardo~D.~Sontag\\
Department of Mathematics\\
Rutgers University, New Brunswick, NJ, USA
}

\maketitle

\begin{abstract}
The multisite phosphorylation-dephosphorylation cycle is a motif repeatedly used in cell signaling. This motif itself can generate a variety of dynamic behaviors like bistability and ultrasensitivity without direct positive feedbacks. In this paper, we study the number of positive steady states of a general multisite phosphorylation-dephosphorylation cycle, and how the number of positive steady states varies by changing the biological parameters. We show analytically that
(1)
for some parameter ranges, there are at least $n+1$ (if $n$ is even) or $n$
(if $n$ is odd) steady states;
(2)
there never are more than $2n-1$ steady states (in particular, this implies
that for $n=2$, including single levels of MAPK cascades, there are at most three steady states); 
(3)
for parameters near the standard Michaelis-Menten quasi-steady
state conditions, there are at most $n+1$ steady states; and
(4)
for parameters far from the standard Michaelis-Menten quasi-steady
state conditions, there is at most one steady state. 
 \end{abstract}


\section{Introduction}
\begin{figure}[h]
  \centering \includegraphics[scale=0.3,angle=0]{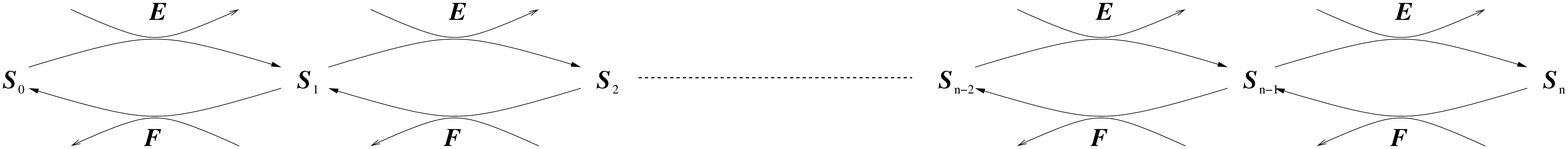}
  \caption{A futile cycle of size $n$.}
\label{fig:Scheme}
\end{figure}

A promising approach to handling the complexity of cell signaling pathways is
to decompose pathways into small motifs, and analyze the individual motifs.
One particular motif that has attracted much attention in recent
years is the cycle formed by two or more inter-convertible forms of one
protein. The protein,
denoted here by $S_0$, is ultimately converted
into a product, denoted here by $S_n$, through a cascade of ``activation''
reactions triggered 
or facilitated by 
an enzyme $E$; conversely, $S_n$ is transformed back (or ``deactivated'')
into the original $S_0$, helped on by the action of a second enzyme $F$.
See Figure \ref{fig:Scheme}.

Such structures, often called ``futile cycles'' (also called substrate cycles, enzymatic
cycles, or enzymatic inter-conversions, see~\cite{Samoilov}), serve as basic
blocks in cellular signaling pathways and have pivotal impact on the signaling
dynamics.   Futile cycles underlie
signaling processes 
such as GTPase cycles \cite{Donovan}, 
bacterial two-component systems and phosphorelays 
\cite{groisman, grossman}
actin treadmilling \cite{chen}),
and glucose mobilization \cite{karp},
as well as metabolic control \cite{stryer}
and cell division and apoptosis \cite{sulis} and cell-cycle checkpoint control
\cite{lew}.
One very important instance is that of Mitogen-Activated Protein Kinase
(``MAPK'') cascades, which regulate primary cellular activities such as
proliferation, differentiation, and 
apoptosis~\cite{lauffenburger,Chang,ferrell,widman}
in eukaryotes from yeast to humans.

MAPK cascades usually consist of three tiers of similar structures with
multiple feedbacks~\cite{Burack,Ferrell_Bhatt,Zhao}. Each individual level of the MAPK cascades is a futile cycle
as depicted in Figure
\ref{fig:Scheme} with $n=2$. 
Markevich {\em et al.}'s paper \cite{Kholodenko} was the first to demonstrate
the possibility of multistationarity at a single cascade level, and
motivated the need for analytical studies of the number of steady
states. Conradi {\em et al.}\ studied the existence of multistationarity in
their paper \cite{conradi}, employing algorithms based on Feinberg's
chemical reaction 
network theory (CRNT). (For more details on CRNT, see \cite{Feinberg,
  Ellison_Feinberg}.) The CRNT algorithm confirms multistationarity in a
single 
level of MAPK cascades, and provides a set of kinetic constants which can give
rise to multistationarity. 
However, the CRNT algorithm
only tests for the existence of multiple steady states, and does not provide
information regarding the precise number of steady states. 

In \cite{jeremy}, Gunawardena proposed a novel approach to the study of steady
states of futile cycles.  His approach, which was focused in the question of
determining the proportion of maximally phosphorylated substrate, was
developed under the simplifying quasi-steady state assumption that substrate
is in excess.  Nonetheless, our study of multistationarity uses in a key
manner the basic formalism in \cite{jeremy}, even for the case when substrate
is not in excess. 

In Section \ref{s:assumption}, we state our basic assumptions regarding the
model. The basic formalism and background for the approach is provided in
Section \ref{s:math}. The main focus of this paper is on Section
\ref{s:bounds}, where we derive various bounds on the number of steady states
of futile cycles of size $n$. The first result is a the lower bound for the
number of steady states. Currently available results on lower bounds, as in
\cite{jeremy2}, can only handle the case when quasi-steady state assumptions
are valid; we substantially extend these results to the fully general case
by means of a perturbation argument which allows one to get around
these restricted assumptions. Another novel feature of our results in
this paper is the derivation of an upper bound of $2n-1$, valid for all kinetic
constants.  Models in molecular cell biology are characterized by a
high degree of uncertainly in parameters, hence such results valid over the
entire parameter space are of special significance.
However, when more information of the parameters are
available, sharper upper bounds can obtained, see Theorems \ref{thm:n+1} and
\ref{thm:1}. We finally conclude our paper in Section \ref{s:end} with a
conjecture of an $n+1$ upper bound.

We remark that the results given here complement our work dealing with the
\emph{dynamical} behavior of futile cycles.  For the case
$n=2$, \cite{IEEEsysbio_WS} showed that the model exhibits generic convergence
to steady states but no more complicated behavior, at least within restricted
parameter
ranges, while \cite{persistencePetri} showed a persistence property (no
species tends to be eliminated) for any possible parameter values.
These papers did not address the question of estimating the \emph{number} of
steady states. 
(An exception is the case $n=1$, for which uniqueness of steady states can be
proved in several ways, and for which global convergence to
these unique equilibria holds \cite{persistencePetri}.)  

\section{Model assumptions}
\label{s:assumption}
Before presenting mathematical details, let us first discuss the basic
biochemical assumptions that go into the model. 
In general, phosphorylation and dephosphorylation can follow either
distributive or processive mechanism. In the processive mechanism, the kinase
(phosphatase) facilitates two or more phosphorylations (dephosphorylations)
before 
the final product is released, whereas in the distributive mechanism, the
kinase (phosphatase) facilitates
at most one phosphorylation (dephosphorylation) in
each molecular encounter. In the case of $n=2$, a futile cycle that follows the
processive mechanism can be represented by reactions as follows:
\begin{align*}
S_0+E &\longleftrightarrow ES_0 \longleftrightarrow ES_1 \longrightarrow S_2+ E \\
S_2+F &\longleftrightarrow FS_2 \longleftrightarrow FS_1 \longrightarrow S_0+ F;
\end{align*}
and the distributive mechanism can be represented by reactions:
\begin{align*}
S_0+E &\longleftrightarrow ES_0 \longrightarrow S_1+E \longleftrightarrow ES_1 \longrightarrow S_2+ E \\
S_2+F &\longleftrightarrow FS_2 \longrightarrow S_1+F \longleftrightarrow FS_1 \longrightarrow S_0+ F.
\end{align*}
Biological experiments have demonstrated that both dual phosphorylation and
dephosphorylation in MAPK are distributive, see
\cite{Burack,Ferrell_Bhatt,Zhao}. In their paper \cite{conradi}, Conradi {\em
  et al.} showed mathematically that if either phosphorylation or
dephosphorylation follows a processive mechanism, the steady state will be
unique,
which, it is argued in \cite{conradi}, contradicts experimental observations.
So, to get more
interesting results, we assume that both phosphorylations and
dephosphorylations in the futile cycles follow the distributive mechanism. 

Our structure of futile cycles in Figure \ref{fig:Scheme} also implicitly
assumes a sequential instead of a random mechanism. By a
sequential mechanism, we mean that the kinase phosphorylates the substrates in
a specific order, and the phosphatase works in the reversed order. This
assumption dramatically reduces the number of different phospho-forms and
simplifies our analysis. In a special case when the kinetic constants of each
phosphorylation are the same and the kinetic constants of each
dephosphorylation are the same, the random mechanism can be easily included in
the sequential case. Biologically, there are systems, for instance the
auto-phosphorylation of FGF-receptor-1, that have been
experimentally shown to follow a sequential mechanism \cite{Furdui}.

To model the reactions, we assume mass action kinetics, which is standard in mathematical modeling of molecular events in biology.

\section{Mathematical formalism}
\label{s:math}
In this section, we set up a mathematical framework for studying the steady states of futile cycles. Let us first write down all the elementary chemical reactions in Figure \ref{fig:Scheme}:
\begin{align*}
S_0+E &\arrowschem{k_{\mbox{on}_0}}{k_{\mbox{off}_0}} ES_0 \stackrel{k_{\mbox{cat}_0}}{\rightarrow} S_1+ E \\
\vdots &   \\
S_{n-1}+E &\arrowschem{k_{\mbox{on}_{n-1}}}{k_{\mbox{off}_{n-1}}}ES_0 \stackrel{k_{\mbox{cat}_{n-1}}}{\rightarrow} S_n+ E \\
\end{align*}
\begin{align*}
S_1+F &\arrowschem{l_{\mbox{on}_0}}{l_{\mbox{off}_0}} FS_1 \stackrel{l_{\mbox{cat}_0}}{\rightarrow} S_0+ F \\
\vdots &   \\
S_n+F &\arrowschem{l_{\mbox{on}_{n-1}}}{l_{\mbox{off}_{n-1}}} FS_n \stackrel{l_{\mbox{cat}_{n-1}}}{\rightarrow} S_{n-1}+ F \\
\end{align*}
where $k_{\mbox{on}_0}$, etc., are kinetic parameters for binding and
unbinding, $ES_0$ denotes the complex consisting of the enzyme $E$ and the
substrate $S_0$, and so forth. These reactions can be modeled by $3n+3$ differential-algebraic equations according to mass action kinetics: 
\begin{align}
\label{eqn:ODE}
\frac{d\s_0}{dt}&=-k_{\mbox{on}_0}\s_0 \e +k_{\mbox{off}_0}\es_0+l_{\mbox{cat}_0}\fs_1  \notag \\
\frac{d\s_i}{dt}&=-k_{\mbox{on}_i}\s_i \e +k_{\mbox{off}_i}\es_i+k_{\mbox{cat}_{i-1}}\es_{i-1}-l_{\mbox{on}_{i-1}}\s_i \f +l_{\mbox{off}_{i-1}}\fs_i+l_{\mbox{cat}_{i}}\fs_{i+1}, \ \ i=1, \dots, n-1 \notag \\
\frac{d \es_j}{dt}&=k_{\mbox{on}_j} \s_j \e-(k_{\mbox{off}_j}+k_{\mbox{cat}_j})\es_j,\ \  j=0,\dots, n-1  \\
\frac{d \fs_k}{dt}&=l_{\mbox{on}_{k-1}}\s_{k} \f-(l_{\mbox{off}_{k-1}}+l_{\mbox{cat}_{k-1}})\fs_k, \ \ k=1, \dots, n, \notag
\end{align}
together with the algebraic ``conservation equations'':
\begin{align}
\label{eqn:conservation}
E_{\mbox{tot}}&=\e+\sum_0^{n-1}\es_i,\notag \\
F_{\mbox{tot}}&=\f+\sum_1^{n}\fs_i, \\
S_{\mbox{tot}}&=\sum_0^n \s_i+\sum_0^{n-1}\es_i+\sum_1^{n}\fs_i.\notag
\end{align}
The variables $\s_0, \dots, \s_n, \es_0,\dots, \es_{n-1}, \fs_1, \dots, \fs_n, e,f$ stand for the concentrations of 
\[
S_0,\dots, S_n, ES_0,\dots, ES_{n-1},FS_1,\dots, FS_n, E,F
\]
respectively. For each positive vector
\begin{align*}
\kappa=&(k_{\mbox{on}_0}, \dots, k_{\mbox{on}_{n-1}}, k_{\mbox{off}_0},\dots, k_{\mbox{off}_{n-1}}, k_{\mbox{cat}_0}, \dots, k_{\mbox{cat}_{n-1}},\\
&l_{\mbox{on}_0}, \dots, l_{\mbox{on}_{n-1}}, l_{\mbox{off}_0},\dots, l_{\mbox{off}_{n-1}}, l_{\mbox{cat}_0}, \dots, l_{\mbox{cat}_{n-1}}) \in \mathbb R^{6n-6}_+
\end{align*}
(of ``kinetic constants'') and each positive triple $\mathcal C=(E_{\mbox{tot}},F_{\mbox{tot}},S_{\mbox{tot}})$, we have a different system $\Sigma(\kappa, \mathcal C)$. 

Let us write the coordinates of a vector $x \in \mathbb R_+^{3n+3}$ as:
\[
x=(\s_0, \dots, \s_n, \es_0,\dots, \es_{n-1}, \fs_1, \dots, \fs_n, e,f),
\]
and define a mapping 
\[
\Phi: \mathbb R_+^{3n+3} \times \mathbb R_+^{6n-6} \times \mathbb R_+^{3} \longrightarrow \mathbb R^{3n+3}
\]
with components $\Phi_1, \dots, \Phi_{3n+3}$ where the first $3n$ components are
\[
\Phi_1(x,\kappa,\mathcal C)=-k_{\mbox{on}_0}\s_0 \e +k_{\mbox{off}_0}\es_0+l_{\mbox{cat}_0}\fs_1,
\]
and so forth, listing the right hand sides of the equations \eqref{eqn:ODE}, $\Phi_{3n+1}$ is
\[
\e+\sum_0^{n-1}\es_i-E_{\mbox{tot}},
\]
and similarly for $\Phi_{3n+2}$ and $\Phi_{3n+3}$, we use the remaining equations in \eqref{eqn:conservation}.

For each $\kappa, \mathcal C$, let us define a set
\[
\mathcal Z(\kappa, \mathcal C)=\{x \, |\, \Phi(x,\kappa,\mathcal C)=0\}.
\]
Observe that, by definition, given $x \in \mathbb R^{3n+3}_+$, $x$ is a positive steady state of $\Sigma(\kappa, \mathcal C)$ if and only if $x \in \mathcal Z(\kappa,\mathcal C)$. So, the mathematical statement of the central problem in this paper is to count the number of elements in $\mathcal Z(\kappa,\mathcal C)$. 
Our analysis will be greatly simplified by a preprocessing. Let us introduce a function
\[
\Psi: \mathbb R_+^{3n+3} \times \mathbb R_+^{6n-6} \times \mathbb R_+^{3} \longrightarrow \mathbb R^{3n+3}
\]
with components $\Psi_1, \dots, \Psi_{3n+3}$ defined as
\begin{align*}
\Psi_1&=\Phi_1+\Phi_{n+1}\\
\Psi_i&=\Phi_i+\Phi_{n+i}+\Phi_{2n+i-1}+\Psi_{i-1},\ \ i=2,\dots,n\\
\Psi_{j}&=\Phi_j, \ \ j=n+1, \dots, 3n+3.
\end{align*}
It is easy to see that
\[
\mathcal Z(\kappa, \mathcal C)=\{x \, |\, \Psi(x,\kappa,\mathcal C)=0\}, 
\]
but now the first $3n$ equations are:
\begin{align*}
\Psi_i&=l_{\mbox{cat}_{i-1}}d_{i}-k_{\mbox{cat}_{i-1}}c_{i-1}=0, \ \ i=1, \dots, n, \\
\Psi_{n+1+j}&=k_{\mbox{on}_j} \s_j \e-(k_{\mbox{off}_j}+k_{\mbox{cat}_j})\es_j=0,\ \  j=0,\dots, n-1  \\
\Psi_{2n+k}&=l_{\mbox{on}_{k-1}}\s_{k} \f-(l_{\mbox{off}_{k-1}}+l_{\mbox{cat}_{k-1}})\fs_k=0, \ \ k=1, \dots, n,
\end{align*}
and can be easily solved as:
\begin{align}
\label{eqn:s_i}
\s_{i+1}&=\lambda_i (e/f)\s_i, \\
\label{eqn:es_i}
\es_i&=\frac{\e\s_i}{K_{M_i}} \\
\label{eqn:fs_i}
\fs_{i+1}&=\frac{\f\s_{i+1}}{L_{M_{i}}},
\end{align}
where 
\begin{equation}
\label{eqn:const}
\lambda_i=\frac{k_{\mbox{cat}_i}L_{M_i}}{K_{M_i}l_{\mbox{cat}_i}}, \ \ K_{M_i}=\frac{k_{\mbox{cat}_i}+k_{\mbox{off}_i}}{k_{\mbox{on}_i}}, \ \ L_{M_i}=\frac{l_{\mbox{cat}_i}+l_{\mbox{off}_i}}{l_{\mbox{on}_i}}, \ \ i =0, \dots, n-1.
\end{equation}
We may now express $\sum_0^n \s_i$, $\sum_0^{n-1} \es_i$ and $\sum_1^n \fs_i$ in terms of $\s_0,\kappa, e$ and $f$:
\begin{align}
\label{eqn:sum}
\sum_{0}^{n}\s_i&=\s_0\left(1+\lambda_0 \left(\frac{e}{f}\right)+\lambda_0\lambda_1 \left(\frac{e}{f}\right)^2+ \cdots +\lambda_0 \cdots \lambda_{n-1} \left(\frac{e}{f}\right)^n\right):=\s_0\varphi_0^\kappa\left(\frac{e}{f}\right), \notag \\
\sum_{0}^{n-1}\es_i&=\e\s_0\left(\frac{1}{K_{M_0}}+\frac{\lambda_0}{K_{M_1}} \left(\frac{e}{f}\right)+\cdots+\frac{\lambda_0\cdots\lambda_{n-2}}{K_{M_{n-1}}} \left(\frac{e}{f}\right)^{n-1}\right):=\e\s_0\varphi_1^\kappa\left(\frac{e}{f}\right),\\
\sum_{1}^{n}\fs_i&=\f\s_0\left(\frac{\lambda_0}{L_{M_0}}\left(\frac{e}{f}\right)+\frac{\lambda_0\lambda_1}{L_{M_1}} \left(\frac{e}{f}\right)^2+\cdots+\frac{\lambda_0\cdots\lambda_{n-1}}{L_{M_{n-1}}} \left(\frac{e}{f}\right)^{n}\right):=\f\s_0\varphi_2^\kappa\left(\frac{e}{f}\right).\notag
\end{align}

Although the equation $\Psi=0$ represents $3n+3$ equations with $3n+3$ unknowns, next we will show that it can be reduced to two equations with two unknowns, which have the same number of positive solutions as $\Psi=0$. Let us first define a set
\[
\mathcal S(\kappa,\mathcal C)=\{(u,v)\in \mathbb R_+ \times \mathbb R_+ \,|\, G_1^{\kappa,\mathcal C}(u,v)=0, G_2^{\kappa,\mathcal C}(u,v)=0\},
\]
where $G_1^{\kappa, \mathcal C}, G_2^{\kappa, \mathcal C}: \mathbb R^2_+ \longrightarrow \mathbb R$ are given by
\begin{align*}
G_1^{\kappa, \mathcal C}(u,v)&=v\left(u\varphi_1^\kappa(u)-\varphi_2^\kappa(u)E_{\mbox{tot}}/F_{\mbox{tot}}\right)-E_{\mbox{tot}}/F_{\mbox{tot}}+u, \\ 
G_2^{\kappa, \mathcal C}(u,v)&=\varphi_0^\kappa(u)\varphi_2^\kappa(u) v^2+\left(\varphi_0^\kappa(u)-S_{\mbox{tot}}\varphi_2^\kappa(u)+F_{\mbox{tot}}u\varphi_1^\kappa(u)+F_{\mbox{tot}}\varphi_2^\kappa(u)\right)v-S_{\mbox{tot}}.
\end{align*}
The precise statement is as follows:
\begin{lemma}
\label{lemma:S}
There exists a mapping $\delta: \mathbb R^{3n+3} \longrightarrow \mathbb R^2$ such that, for each $\kappa, \mathcal C$, the map $\delta$ restricted to $\mathcal Z(\kappa, \mathcal C)$ is a bijection between the sets $\mathcal Z(\kappa, \mathcal C)$ and $\mathcal S(\kappa, \mathcal C)$.
\end{lemma}
\begin{proof}
Let us define the mapping $\delta: \mathbb R^{3n+3} \longrightarrow \mathbb R^2$ as $\delta(x)=(e/f,\s_0)$, where 
\[
x=(\s_0, \dots, \s_n, \es_0, \dots, \es_{n-1}, \fs_1, \dots, \fs_n, e,f).
\]
If we can show that $\delta$ induces a bijection between $\mathcal Z(\kappa, \mathcal C)$ and $\mathcal S(\kappa, \mathcal C)$, we are done.

First, we claim that $\delta(\mathcal Z(\kappa, \mathcal C))\subseteq \mathcal S(\kappa, \mathcal C)$. Pick any $x \in \mathcal Z(\kappa, \mathcal C)$, we have that $x$ satisfies \eqref{eqn:s_i}-\eqref{eqn:fs_i}. 
Moreover, $\Phi_{3n+2}(x,\kappa, \mathcal C)=0$ yields 
\[
E_{\mbox{tot}}=e+e\s_0\varphi_1^\kappa(\frac{e}{f}),
\]
and thus 
\begin{equation}
\label{eqn:e}
e=\frac{E_{\mbox{tot}}}{1+\s_0 \varphi_1^\kappa(e/f)}.
\end{equation}
Using $\Phi_{3n+1}(x,\kappa, \mathcal C)=0$ and $\Phi_{3n+2}(x,\kappa, \mathcal C)=0$, we get:
\begin{equation}
\label{eqn:ratio}
\frac{E_{\mbox{tot}}}{F_{\mbox{tot}}}=\frac{e(1+\s_0\varphi_1^\kappa(e/f))}{f(1+\s_0\varphi_2^\kappa(e/f))},
\end{equation}
which is $G_1^{\kappa, \mathcal C}(e/f,\s_0)=0$ after multiplying by $1+\s_0\varphi_2^\kappa(e/f)$ and rearranging terms. 

To check that $G_2^{\kappa, \mathcal C}(e/f,\s_0)=0$, we start with $\Phi_{3n+3}(x,\kappa, \mathcal C)=0$, i.e.
\[
S_{\mbox{tot}}=\sum_0^n \s_i+\sum_0^{n-1}\es_i+\sum_1^n\fs_i.
\]
Using \eqref{eqn:sum} and \eqref{eqn:e}, this expression becomes
\begin{align*}
S_{\mbox{tot}}&=\s_0\varphi_0^\kappa(\frac{e}{f})+\frac{E_{\mbox{tot}}\s_0\varphi_1^\kappa(e/f)}{1+\s_0\varphi_1^\kappa(e/f)}+\frac{F_{\mbox{tot}}\s_0\varphi_2^\kappa(e/f)}{1+\s_0\varphi_2^\kappa(e/f)}\\
&=\s_0\varphi_0^\kappa(\frac{e}{f})+\frac{eF_{\mbox{tot}}\s_0\varphi_1^\kappa(e/f)}{f(1+\s_0\varphi_2^\kappa(e/f))}+\frac{F_{\mbox{tot}}\s_0\varphi_2^\kappa(e/f)}{1+\s_0\varphi_2^\kappa(e/f)},
\end{align*}
where the last equality comes from \eqref{eqn:ratio}.

After multiplying by $1+\s_0\varphi_2^\kappa(e/f)$, and simplifying, we get
\[
\varphi_0^\kappa(\frac{e}{f})\varphi_2^\kappa(\frac{e}{f}) \s_0^2+\left(\varphi_0^\kappa(\frac{e}{f})-S_{\mbox{tot}}\varphi_2^\kappa(\frac{e}{f})+\frac{e}{f}F_{\mbox{tot}}\varphi_1^\kappa(\frac{e}{f})+F_{\mbox{tot}}\varphi_2^\kappa(u)\right)\s_0-S_{\mbox{tot}}=0,
\]
that is, $G_2^{\kappa,\mathcal C}(e/f,\s_0)=0$. since both $G_1^{\kappa,\mathcal C}(e/f,\s_0)$ and $G_2^{\kappa,\mathcal C}(e/f,\s_0)$ are zero,  $\delta(x) \in \mathcal S(\kappa,\mathcal C)$. 

Next, we will show that $\mathcal S(\kappa, \mathcal C) \subseteq \delta(\mathcal Z(\kappa, \mathcal C))$. For any $y=(u,v) \in \mathcal S(\kappa,\mathcal C)$, let the coordinates of $x$ be defined as:
\begin{align*}
\s_0&=v\\
\s_{i+1}&=\lambda_i u\s_i \notag \\
\e&=\frac{E_{\mbox{tot}}}{1+\s_0\varphi_1^\kappa(u)} \notag \\
\f&= \frac{\e}{u} \\
\es_i&=\frac{\e\s_i}{K_{M_i}} \notag \\
\fs_{i+1}&=\frac{\f\s_{i+1}}{L_{M_{i}}} \notag
\end{align*}
for $i=0, \dots, n-1$. It is easy to see that the vector $x=(\s_0, \dots, \s_n, \es_0, \dots, \es_{n-1}, \fs_1, \dots, \fs_n, e,f)$ satisfies $\Phi_{1}(x,\kappa, \mathcal C)=0,\dots,\Phi_{3n+1}(x,\kappa, \mathcal C)=0$. If $\Phi_{3n+2}(x,\kappa, \mathcal C)$ and $\Phi_{3n+3}(x,\kappa, \mathcal C)$ are also zero, then $x$ is an element of $\mathcal Z(\kappa, \mathcal C)$ with $\delta(x)=y$. Given the condition that $G_i^{\kappa, \mathcal C}(u,v)=0$ ($i=1,2$) and $u=e/f, v=\s_0$, we have $G_1^{\kappa,\mathcal C}(e/f,\s_0)=0$, and therefore \eqref{eqn:ratio} holds.
Since
\[
\e=\frac{E_{\mbox{tot}}}{1+\s_0\varphi_1^\kappa(e/f)}
\]
in our construction, we have
\[
F_{\mbox{tot}}=f(1+\s_0\varphi_2^\kappa(e/f))=f+\sum_1^n\fs_i.
\]
To check $\Phi_{3n+3}(x,\kappa, \mathcal C)=0$, we use 
\[
\frac{G_2^{\kappa,\mathcal C}(e/f,\s_0)}{1+\s_0\varphi_2^\kappa(e/f)}=0,
\]
as $G_2^{\kappa,\mathcal C}(e/f,\s_0)=0$ and $1+\s_0\varphi_2^\kappa(e/f)>0$. Applying \eqref{eqn:sum}-\eqref{eqn:ratio}, we have
\[
\sum_0^n \s_i+\sum_0^{n-1} \es_i +\sum_1^n \fs_i=\s_0\varphi_0^\kappa(e/f)+\frac{eF_{\mbox{tot}}\s_0\varphi_1^\kappa(e/f)}{f(1+\s_0\varphi_2^\kappa(e/f))}+\frac{F_{\mbox{tot}}\s_0\varphi_2^\kappa(e/f)}{1+\s_0\varphi_2^\kappa(e/f)}=S_{\mbox{tot}}.
\]
It remains for us to show that the map $\delta$ is one to one on $\mathcal Z(\kappa, \mathcal C)$. Suppose that $\delta(x^1)=\delta(x^2)=(u,v)$, where 
\[
x^i=(\s_0^i, \dots, \s_{n}^i, \es_0^i, \dots, \es_{n-1}^i, \fs_1^i,\dots, \fs_n^i, e^i,f^i), \ \ i=1,2.
\] 
By the definition of $\delta$, we know that $\s_0^1=\s_0^2$ and $e^1/f^1=e^2/f^2$. Therefore, $\s_i^1=\s_i^2$ for $i=0,\dots, n$. Equation \eqref{eqn:e} gives
\[
e^1=\frac{E_{\mbox{tot}}}{1+v\varphi_1^\kappa(u)}=e^2.
\]
Thus, $f^1=f^2$, and $\es_i^1=\es_i^2,\fs_{i+1}^1=\fs_{i+1}^2$ for $i=0,\dots, n-1$ because of \eqref{eqn:s_i}-\eqref{eqn:fs_i}. Therefore, $x^1=x^2$, and $\delta$ is one to one.
\end{proof}

The above lemma ensures that the two sets $\mathcal Z(\kappa,\mathcal C)$ and $\mathcal S(\kappa,\mathcal C)$ have the same number of elements. From now on, we will focus on $\mathcal S(\kappa,\mathcal C)$, the set of positive solutions of equations $G_1^{\kappa,\mathcal C}(u,v)=0, G_2^{\kappa,\mathcal C}(u,v)=0$, i.e.
\begin{align}
\label{eqn:G1}
G_1^{\kappa, \mathcal C}(u,v)&=v\left(u\varphi_1^\kappa(u)-\varphi_2^\kappa(u)E_{\mbox{tot}}/F_{\mbox{tot}}\right)-E_{\mbox{tot}}/F_{\mbox{tot}}+u=0, \\ 
\label{eqn:G2}
G_2^{\kappa, \mathcal C}(u,v)&=\varphi_0^\kappa(u)\varphi_2^\kappa(u) v^2+\left(\varphi_0^\kappa(u)-S_{\mbox{tot}}\varphi_2^\kappa(u)+F_{\mbox{tot}}u\varphi_1^\kappa(u)+F_{\mbox{tot}}\varphi_2^\kappa(u)\right)v-S_{\mbox{tot}}=0.
\end{align}
\section{Number of positive steady states}
\label{s:bounds}

\subsection{Lower bound on the number of positive steady states}

One approach to solving \eqref{eqn:G1}-\eqref{eqn:G2} is to view $G_2^{\kappa,\mathcal C}(u,v)$ as a quadratic polynomial in $v$.
Since $G_2^{\kappa, \mathcal C}(u,0)<0$, equation \eqref{eqn:G2} has a unique positive root, namely
\begin{equation}
\label{eqn:v}
v=\frac{-H^{\kappa, \mathcal C}(u)+\sqrt{H^{\kappa, \mathcal C}(u)^2+4S_{\mbox{tot}}\varphi_0^\kappa(u)\varphi_2^\kappa(u)}}{2\varphi_0^\kappa(u)\varphi_2^\kappa(u)},
\end{equation}
where
\begin{equation}
\label{eqn:H}
H^{\kappa, \mathcal C}(u)=\varphi_0^\kappa(u)-S_{\mbox{tot}}\varphi_2^\kappa(u)+F_{\mbox{tot}}u\varphi_1^\kappa(u)+F_{\mbox{tot}}\varphi_2^\kappa(u).
\end{equation}
Substituting this expression for $v$ into \eqref{eqn:G1}, and multiplying by $\varphi_0^\kappa(u)$, we get
\begin{equation}
\label{eqn:Fu}
F^{\kappa, \mathcal C}(u):=\frac{-\tilde H^{\kappa, \mathcal C}(u)+\sqrt{\tilde H^{\kappa, \mathcal C}(u)^2+4S_{\mbox{tot}}\varphi_0^\kappa(u)\varphi_2^\kappa(u)}}{2\varphi_2^\kappa(u)}\left(u\varphi_1^\kappa(u)-\frac{E_{\mbox{tot}}}{F_{\mbox{tot}}}\varphi_2^\kappa(u)\right)-\frac{E_{\mbox{tot}}}{F_{\mbox{tot}}}\varphi_0^\kappa(u)+u\varphi_0^\kappa(u)=0.
\end{equation}
So, any $(u,v) \in \mathcal S(\kappa, \mathcal C)$ should satisfy \eqref{eqn:v} and \eqref{eqn:Fu}. On the other hand, any positive solution $u$ of \eqref{eqn:Fu} (notice that $\varphi_0^\kappa(u)>0$) and $v$ given by \eqref{eqn:v} (always positive) provide a positive a solution of \eqref{eqn:G1}-\eqref{eqn:G2}, that is, $(u,v)$ is an element in $\mathcal S(\kappa, \mathcal C)$. Therefore, the number of positive solutions of \eqref{eqn:G1}-\eqref{eqn:G2} is the same as the number of positive solutions of \eqref{eqn:v} and \eqref{eqn:Fu}. But $v$ is uniquely determined by $u$ in \eqref{eqn:v}, which further simplifies the problem to one equation \eqref{eqn:Fu} with one unknown $u$. Based on this observation, we have:
\begin{theorem}
\label{thm:exist}
For each positive numbers $S_{\mbox{tot}},\gamma$, there exist $\varepsilon_0>0$ and $\kappa \in \mathbb R^{6n-6}_+$ such that the following property holds. Pick any $E_{\mbox{tot}}, F_{\mbox{tot}}$ such that 
\begin{equation}
\label{eqn:EFS}
F_{\mbox{tot}}=E_{\mbox{tot}}/\gamma< \varepsilon_0 S_{\mbox{tot}}/\gamma,
\end{equation}
then the system $\Sigma(\kappa,\mathcal C)$ with $\mathcal C=(E_{\mbox{tot}}, F_{\mbox{tot}}, S_{\mbox{tot}})$ has at least $n+1$ ($n$) positive steady states when $n$ is even (odd).
\end{theorem}
\begin{proof}
For each $\kappa, \gamma, S_{\mbox{tot}}$, let us define two functions $\mathbb R_+ \times \mathbb R_+ \longrightarrow \mathbb R$ as follows:
\begin{align}
\label{eqn:H1}
\tilde H^{\kappa, \gamma, S_{\mbox{tot}}}(\varepsilon,u)&=H^{\kappa, (\varepsilon S_{\mbox{tot}}, \varepsilon S_{\mbox{tot}}/\gamma, S_{\mbox{tot}})}(u)\\
&=\varphi_0^\kappa(u)-S_{\mbox{tot}}\varphi_2^\kappa(u)+\varepsilon\frac{S_{\mbox{tot}}}{\gamma}u\varphi_1^\kappa(u)+\varepsilon\frac{S_{\mbox{tot}}}{\gamma}\varphi_2^\kappa(u), \notag
\end{align}
and
\begin{align}
\label{eqn:F}
\tilde F^{\kappa, \gamma, S_{\mbox{tot}}}(\varepsilon,u)&=F^{\kappa, (\varepsilon S_{\mbox{tot}}, \varepsilon S_{\mbox{tot}}/\gamma, S_{\mbox{tot}})}(u)\\
&=\frac{-\tilde H^{\kappa, \gamma, S_{\mbox{tot}}}(\varepsilon,u)+\sqrt{\tilde H^{\kappa, \gamma, S_{\mbox{tot}}}(\varepsilon,u)^2+4S_{\mbox{tot}}\varphi_0^\kappa(u)\varphi_2^\kappa(u)}}{2\varphi_2^\kappa(u)}\left(u\varphi_1^\kappa(u)-\gamma\varphi_2^\kappa(u)\right) \notag \\
&-\gamma\varphi_0^\kappa(u)+u\varphi_0^\kappa(u).\notag
\end{align}
By Lemma \ref{lemma:S} and the argument before this theorem, it is enough to show that there exist $\varepsilon_0>0$ and $\kappa \in \mathbb R^{6n-6}_+$ such that for all $\varepsilon \in (0,\varepsilon_0)$, the equation $\tilde F^{\kappa, \gamma, S_{\mbox{tot}}}(\varepsilon,u)=0$ has at least $n+1$ ($n$) positive solutions when $n$ is even (odd). (Then, given $S_{\mbox{tot}}$, $\gamma$, $E_{\mbox{tot}}$, and $F_{\mbox{tot}}$ satisfying
\eqref{eqn:EFS}, we let $\varepsilon = E_{\mbox{tot}}/S_{\mbox{tot}} < \varepsilon_0$, and apply the result.)

A straightforward computation shows that when $\varepsilon=0$, 
\begin{align}
\label{eqn:f}
\tilde F^{\kappa, \gamma, S_{\mbox{tot}}}(0,u)&=S_{\mbox{tot}}\left(u\varphi_1^\kappa(u)-\gamma \varphi_2^\kappa(u)\right)-\gamma \varphi_0^\kappa(u)+u\varphi_0^\kappa(u) \notag \\
&=\lambda_0 \cdots \lambda_{n-1} u^{n+1} +\lambda_0 \cdots \lambda_{n-2}\left(1+\frac{S_{\mbox{tot}}}{K_{M_{n-1}}}\left(1-\gamma \beta_{n-1}\right)-\gamma \lambda_{n-1} \right) u^n \notag \\
&+\cdots +\lambda_0 \cdots \lambda_{i-2}\left(1+\frac{S_{\mbox{tot}}}{K_{M_{i-1}}}\left(1-\gamma \beta_{i-1}\right)-\gamma \lambda_{i-1} \right) u^i+\cdots \\
&+\left(1+\frac{S_{\mbox{tot}}}{K_{M_{0}}}\left(1-\gamma \beta_{0}\right)-\gamma \lambda_{0} \right) u -\gamma, \notag
\end{align}
where the $\lambda_i$'s and $K_{M_i}$'s are defined as in \eqref{eqn:const}, and $\beta_i=k_{\mbox{cat}_i}/l_{\mbox{cat}_i}$. The polynomial $\tilde F^{\kappa, \gamma, S_{\mbox{tot}}}(0,u)$ is of degree $n+1$, so there are at most $n+1$ positive roots. Notice that $u=0$ is not a root because $\tilde F^{\kappa, \gamma, S_{\mbox{tot}}}(0,u)=-\gamma<0$, which also implies that when $n$ is odd, there can not be $n+1$ positive roots. Now fix any $S_{\mbox{tot}}$ and $\gamma$. We will construct a vector $\kappa$ such that $\tilde F^{\kappa, \gamma, S_{\mbox{tot}}}(0,u)$ has $n+1$ distinct positive roots when $n$ is even.

Let us pick any $n+1$ positive real numbers $u_1< \cdots< u_{n+1}$, such that their product is $\gamma$, and assume that 
\begin{equation}
\label{eqn:u}
(u-u_1)\cdots(u-u_{n+1})=u^{n+1}+a_n u^n+\cdots + a_1 u+a_0,
\end{equation}
where $a_0=-\gamma<0$ keeping in mind that $a_i$'s are given. Our goal is to find a vector $\kappa \in \mathbb R_+^{6n-6}$ such that \eqref{eqn:f} and \eqref{eqn:u} are the same. For each $i=0,\dots, n-1$, we pick $\lambda_i=1$. Comparing the coefficients of $u^{i+1}$ in \eqref{eqn:f} and \eqref{eqn:u}, we have:
\begin{equation}
\label{eqn:compare}
\frac{S_{\mbox{tot}}}{K_{M_{i}}}(1+a_0\beta_{i})=a_{i+1}-a_0-1.
\end{equation}
Let us pick $K_{M_i}>0$
such that $\frac{K_{M_i}}{S_{\mbox{tot}}}(a_{i+1}-a_0-1)-1<0$, then take
\[
\beta_i=\frac{\frac{K_{M_i}}{S_{\mbox{tot}}}(a_{i+1}-a_0-1)-1}{a_0}>0
\]
in order to satisfy \eqref{eqn:compare}. From the given
\[
\lambda_0, \dots, \lambda_{n-1}, K_{M_0},\dots, K_{M_{n-1}}, \beta_0,\dots, \beta_{n-1},
\]
we will find a vector
\begin{align*}
\kappa=&(k_{\mbox{on}_0}, \dots, k_{\mbox{on}_{n-1}}, k_{\mbox{off}_0},\dots, k_{\mbox{off}_{n-1}}, k_{\mbox{cat}_0}, \dots, k_{\mbox{cat}_{n-1}},\\
&l_{\mbox{on}_0}, \dots, l_{\mbox{on}_{n-1}}, l_{\mbox{off}_0},\dots, l_{\mbox{off}_{n-1}}, l_{\mbox{cat}_0}, \dots, l_{\mbox{cat}_{n-1}}) \in \mathbb R^{6n-6}_+
\end{align*}
such that $\beta_i=k_{\mbox{cat}_i}/l_{\mbox{cat}_i}, i=0,\dots, n-1,$ and \eqref{eqn:const} holds.
This vector $\kappa$ will guarantee that $\tilde F^{\kappa, \gamma, S_{\mbox{tot}}}(0,u)$ has $n+1$ positive distinct roots. When $n$ is odd, a similar construction will give a vector $\kappa$ such that $\tilde F^{\kappa, \gamma, S_{\mbox{tot}}}(0,u)$ has $n$ positive roots and one negative root.

One construction of $\kappa$ (given $\lambda_i, K_{M_i}, \beta_i, i=0,\dots, n-1$) is as follows. For each $i=0, \dots, n-1$, we start by defining:
 \[
 L_{M_i}=\frac{\lambda_i K_{M_i}}{\beta_i},
 \]
consistently with the definitions in \eqref{eqn:const}. Then, we take 
\[
k_{\mbox{on}_i}=1, \ \ l_{\mbox{on}_i}=1,
\]
and 
\[
k_{\mbox{off}_i}=\alpha_i K_{M_i}, \ \ k_{\mbox{cat}_i}=(1-\alpha_i) K_{M_i}, \ \ l_{\mbox{cat}_i}=\frac{1-\alpha_i}{\beta_i} K_{M_i}, \ \ l_{\mbox{off}_i}=L_{M_i}-l_{\mbox{cat}_i},
\]
where $\alpha_i \in (0,1)$ is chosen such that 
\[
l_{\mbox{off}_i}= L_{M_i}-\frac{1-\alpha_i}{\beta_i}K_{M_i}>0.
\]
This $\kappa$ satisfies $\beta_i=k_{\mbox{cat}_i}/l_{\mbox{cat}_i}, i=0,\dots, n-1,$ and \eqref{eqn:const}.

In order to apply the Implicit Function Theorem, we now view the functions defined by formulas in \eqref{eqn:H1} and \eqref{eqn:F} as defined also for $\varepsilon \leq 0$, i.e. as functions $\mathbb R\times \mathbb R_+ \longrightarrow \mathbb R$.
It is easy to see that $\tilde F^{\kappa, \gamma, S_{\mbox{tot}}}(\varepsilon,u)$ is $C^1$ on $\mathbb R \times \mathbb R_+$ because the polynomial under the square root sign in $\tilde F^{\kappa, \gamma, S_{\mbox{tot}}}(\varepsilon,u)$ is never zero.
On the other hand, since $\tilde F^{\kappa, \gamma, S_{\mbox{tot}}}(0,u)$ is a polynomial in $u$ with distinct roots, $\frac{\partial \tilde F^{\kappa, \gamma, S_{\mbox{tot}}}}{\partial u}(0,u_i) \not=0$. By the Implicit Function Theorem, for each $i=1,\dots, n+1$, there exist open intervals $E_i$ containing $0$, and open intervals $U_i$ containing $u_i$, and a differentiable function 
\[
\alpha_i: E_i \rightarrow U_i
\]
such that $\alpha_i(0)=u_i$, $\tilde F^{\kappa, \gamma, S_{\mbox{tot}}}(\varepsilon, \alpha_i(\varepsilon))=0$ for all $\varepsilon \in E_i$, and the images $\alpha_i(E_i)$'s are non-overlapping. If we take
\[
(0,\varepsilon_0):=\bigcap_{1}^{n+1}E_i \bigcap \ (0,+\infty),
\]
then for any $\varepsilon \in (0,\varepsilon_0)$, we have $\{\alpha_i(\varepsilon)\}$ as $n+1$ distinct positive roots of $\tilde F^{\kappa, \gamma, S_{\mbox{tot}}}(\varepsilon, u)$. The case when $n$ is odd can be proved similarly.
\end{proof}

\begin{figure}[h]
  \centering \includegraphics[scale=0.3,angle=0]{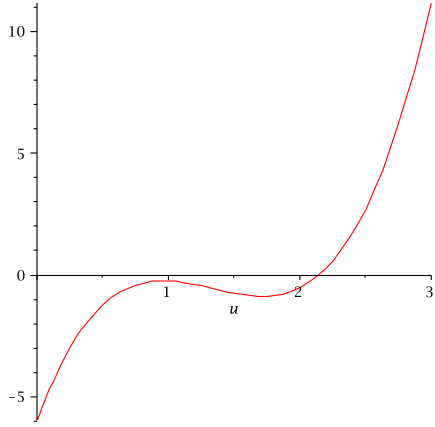}
  \caption{The plot of the function $\tilde F^{\kappa, \gamma, S_{\mbox{tot}}}(0.01,u)$ on $[0,3]$. There is a unique positive real solution around $u=2.14$, the double root $u=1$ of $\tilde F^{\kappa, \gamma, S_{\mbox{tot}}}(0,u)$ bifurcates to two complex roots with non-zero imaginary parts.}
\label{fig:counter}
\end{figure}

The above theorem shows that when $E_{\mbox{tot}}/F_{\mbox{tot}}$ is sufficiently small, it is always possible for the futile cycle to have $n+1$ ($n$) steady states when $n$ is even (odd), by choosing appropriate kinetic constants $\kappa$. We should notice that for arbitrary $\kappa$, the derivative of $\tilde {\mathcal F}$ at each positive root may become zero, which breaks down the perturbation argument. Here is an example to show that more conditions are needed: with
\[
n=2,\ \ \lambda_0=1,\ \ \lambda_1=3,\ \ \gamma=6,\ \  \beta_0=\beta_1=1/12,\ \  K_0=1/8,\ \  K_1=1/2, \ \ S_{\mbox{tot}}=5,
\]
we have that
\[
\tilde F^{\kappa, \gamma, S_{\mbox{tot}}}(0,u)=3u^3-12u^2+15u-6=3(u-1)^2(u-2)
\]
has a double root at $u=1$. In this case, even for $\varepsilon=0.01$, there is only one positive root of $\tilde F^{\kappa, \gamma, S_{\mbox{tot}}}(\varepsilon,u)$, see Figure \ref{fig:counter}.

However, the following lemma provides a sufficient condition for $\frac{\partial F^{\kappa, \gamma, S_{\mbox{tot}}}}{\partial u}(0,\bar u) \not=0$, for any positive $\bar u$ such that $\tilde F^{\kappa, \gamma, S_{\mbox{tot}}}(0,\bar u)=0$.
\begin{lemma}
\label{lemma:derivative}
For each positive numbers $S_{\mbox{tot}},\gamma$, and vector $\kappa \in \mathbb R^{6n-6}_+$, if
\begin{equation}
\label{eqn:condition}
S_{\mbox{tot}}\left|\frac{1-\gamma \beta_{j}}{K_{M_{j}}}\right|\leq \frac{1}{n}
\end{equation}
holds for all $j=1, \cdots, n-1$, then $\frac{\partial \tilde F^{\kappa, \gamma, S_{\mbox{tot}}}}{\partial u}(0,\bar u) \not=0$.
\end{lemma}

See Appendix for the proof.

\begin{theorem}
For each positive numbers $S_{\mbox{tot}}, \gamma$, and vector $\kappa \in \mathbb R^{6n-6}_+$ satisfying condition \eqref{eqn:condition}, there exists $\varepsilon_1>0$ such that for any $F_{\mbox{tot}}, E_{\mbox{tot}}$ satisfying $F_{\mbox{tot}}= E_{\mbox{tot}}/\gamma< \varepsilon_1 S_{\mbox{tot}}/\gamma$, the number of positive steady states of system $\Sigma(\kappa, \mathcal C)$ is greater or equal to the number of (positive) roots of $\tilde F^{\kappa, \gamma, S_{\mbox{tot}}}(0, u)$.
\end{theorem}
\begin{proof}
Suppose that $\tilde F^{\kappa, \gamma, S_{\mbox{tot}}}(0, u)$ has $m$ roots: $\bar u_1, \dots, \bar u_m$. Applying Lemma \ref{lemma:derivative}, we have 
\[
\frac{\partial \tilde F^{\kappa, \gamma, S_{\mbox{tot}}}}{\partial u}(0,\bar u_k) \not=0, k=1,\dots, m.
\]
By the perturbation arguments as in Theorem \ref{thm:exist}, we have that there exists $\varepsilon_1>0$ such that $\tilde F^{\kappa, \gamma, S_{\mbox{tot}}}(\varepsilon, u)$ has at least $m$ roots for all $0<\varepsilon<\varepsilon_1$.
\end{proof}

The above result depends heavily on a perturbation argument, which only works when $E_{\mbox{tot}}/F_{\mbox{tot}}$ is sufficiently small. In the next section, we will give an upper bound of the number of steady states with no restrictions on $E_{\mbox{tot}}/F_{\mbox{tot}}$, and independent of $\kappa$ and $\mathcal C$.
\subsection{Upper bound on the number of steady states}
\begin{theorem}
\label{thm:upper_bound}
For each $\kappa, \mathcal C$, the system $\Sigma(\kappa, \mathcal C)$ has at most $2n-1$ positive steady states.
\end{theorem}
\begin{proof}
An alternative approach to solving \eqref{eqn:G1}-\eqref{eqn:G2} is to first eliminate $v$ from \eqref{eqn:G1} instead of from \eqref{eqn:G2}, i.e.
\newcommand{\EoF}{(E_{\mbox{tot}}/F_{\mbox{tot}})} 
\begin{equation}
\label{eqn:vv}
v=\frac{E_{\mbox{tot}}/F_{\mbox{tot}}-u}{u\varphi_1^\kappa(u)-\EoF\varphi_2^\kappa(u)}:=\frac{A(u)}{B(u)},
\end{equation}
when $u\varphi_1^\kappa(u)-\EoF\varphi_2^\kappa(u) \not= 0$. Then, we substitute \eqref{eqn:vv} into \eqref{eqn:G2}, and multiply by $(u\varphi_1^\kappa(u)-\EoF\varphi_2^\kappa(u))^2$ to get:
\begin{align}
\label{eqn:uu}
P^{\kappa, \mathcal C}(u)&:=\varphi_0^\kappa\varphi_2^\kappa\left(\frac{E_{\mbox{tot}}}{F_{\mbox{tot}}}-u\right)^2+\left( \varphi_0^\kappa-S_{\mbox{tot}}\varphi_2^\kappa+F_{\mbox{tot}}u\varphi_1^\kappa+F_{\mbox{tot}}\varphi_2^\kappa \right)\left(\frac{E_{\mbox{tot}}}{F_{\mbox{tot}}}-u\right)\left(u\varphi_1^\kappa-\frac{E_{\mbox{tot}}}{F_{\mbox{tot}}} \varphi_2^\kappa\right) \notag \\
&-S_{\mbox{tot}}\left(u\varphi_1^\kappa-\frac{E_{\mbox{tot}}}{F_{\mbox{tot}}} \varphi_2^\kappa\right)^2=0.
\end{align}
Therefore, if $u\varphi_1^\kappa(u)-\EoF\varphi_2^\kappa(u) \not= 0$, the number of positive solutions of \eqref{eqn:G1}-\eqref{eqn:G2} is no greater than the number of positive roots of $P^{\kappa, \mathcal C}(u)$.

In the special case when $u\varphi_1^\kappa(u)-\EoF\varphi_2^\kappa(u)=0$, by \eqref{eqn:G1}, we must have $u=E_{\mbox{tot}}/F_{\mbox{tot}}$, and thus $\varphi_1^\kappa(E_{\mbox{tot}}/F_{\mbox{tot}})=\varphi_2^\kappa(E_{\mbox{tot}}/F_{\mbox{tot}})$. Substituting into \eqref{eqn:G2}, we get a unique $v$ defined as in \eqref{eqn:v} with $u=E_{\mbox{tot}}/F_{\mbox{tot}}$. But notice that in this case $u=E_{\mbox{tot}}/F_{\mbox{tot}}$ is also a root of $P^{\kappa, \mathcal C}(u)$, so also in this case the number of positive solutions to \eqref{eqn:G1}-\eqref{eqn:G2} is no greater than the number of positive roots of $P^{\kappa, \mathcal C}(u)$. 

It is easy to see that $P^{\kappa, \mathcal C}(u)$ is divisible by $u$. Consider the polynomial $Q^{\kappa, \mathcal C}(u):=P^{\kappa, \mathcal C}(u)/u$ of degree $2n+1$. We will first show that $Q^{\kappa, \mathcal C}(u)$ has no more than $2n$ positive roots, then we will prove by contradiction that $2n$ distinct positive roots can not be achieved.

It is easy to see that the coefficient of $u^{2n+1}$ is 
\[
\frac{(\lambda_0  \cdots \lambda_{n-1})^2}{L_{M_{n-1}}}>0,
\]
and the constant term is 
\[
\frac{E_{\mbox{tot}}}{F_{\mbox{tot}}K_{M_0}}>0.
\]
So the polynomial $Q^{\kappa, \mathcal C}(u)$ has at least one negative root, and thus has no more than $2n$ positive roots.

Suppose that $\mathcal S(\kappa, \mathcal C)$ has cardinality $2n$, then $Q^{\kappa, \mathcal C}(u)$ must have $2n$ distinct positive roots, and each of them has multiplicity one. Let us denote the roots as $u_1,\dots,u_{2n}$ in ascending order. We claim that none of them equals $E_{\mbox{tot}}/F_{\mbox{tot}}$. If so, we would have $\varphi_1^\kappa(E_{\mbox{tot}}/F_{\mbox{tot}})=\varphi_2^\kappa(E_{\mbox{tot}}/F_{\mbox{tot}})$, and $E_{\mbox{tot}}/F_{\mbox{tot}}$ would be a double root of $Q^{\kappa, \mathcal C}(u)$, contradiction. 

Since $Q^{\kappa, \mathcal C}(0)>0$, $Q^{\kappa, \mathcal C}(u)$ is positive on intervals 
\[
I_0=(0,u_1), I_1=(u_2,u_3), \dots, I_{n-1}=(u_{2n-2},u_{2n-1}),I_n=(u_{2n},\infty),
\]
and negative on intervals
\[
J_1=(u_1,u_2), \dots, J_{n}=(u_{2n-1},u_{2n}).
\]
As remarked earlier,  $\varphi_1^\kappa(E_{\mbox{tot}}/F_{\mbox{tot}})\not=\varphi_2^\kappa(E_{\mbox{tot}}/F_{\mbox{tot}})$, the polynomial $Q^{\kappa, \mathcal C}(u)$ evaluated at $E_{\mbox{tot}}/F_{\mbox{tot}}$ is negative, and therefore, $E_{\mbox{tot}}/F_{\mbox{tot}}$ belongs to one of the $J$ intervals, say $J_s=(u_{2s-1}, u_{2s})$, for some $s \in \{1,\dots, n\}$ .

On the other hand, the denominator of $v$ in \eqref{eqn:vv}, denoted as $B(u)$, is a polynomial of degree $n$ and divisible by $u$. If $B(u)$ has no positive root, then it does not change sign on the positive axis of $u$. But $v$ changes sign when $u$ passes $E_{\mbox{tot}}/F_{\mbox{tot}}$, thus $v_{2s-1}$ and $v_{2s}$ have opposite signs, and one of $(u_{2s-1}, v_{2s-1})$ and $(u_{2s}, v_{2s})$ is not a solution to \eqref{eqn:G1}-\eqref{eqn:G2}, which contradicts the fact that both are in $\mathcal S(\kappa, \mathcal C)$.

Otherwise, there exists a positive root $\bar u$ of $B(u)$ such that there is no other positive root of $B(u)$ between $\bar u$ and $E_{\mbox{tot}}/F_{\mbox{tot}}$. Plugging $\bar u$ into $Q^{\kappa, \mathcal C}(u)$, we see that $Q^{\kappa, \mathcal C}(\bar u)$ is always positive, therefore, $\bar u$ belongs to one of the $I$ intervals, say $I_{t}=(u_{2t}, u_{2t+1})$ for some $t \in \{0,\dots, n\}$. There are two cases:
\begin{enumerate}
\item $E_{\mbox{tot}}/F_{\mbox{tot}}<\bar u$. We have 
\[
u_{2s-1}<E_{\mbox{tot}}/F_{\mbox{tot}}<u_{2t}<\bar u.
\]
Notice that $v$ changes sign when $u$ passes $E_{\mbox{tot}}/F_{\mbox{tot}}$, so the corresponding $v_{2s-1}$ and $v_{2t}$ have different signs, and either $(u_{2s-1},v_{2s-1}) \notin \mathcal S(\kappa, \mathcal C)$ or $(u_{2t},v_{2t}) \notin \mathcal S(\kappa, \mathcal C)$, contradiction.
\item $E_{\mbox{tot}}/F_{\mbox{tot}}>\bar u$. We have
\[
\bar u<u_{2t+1}<E_{\mbox{tot}}/F_{\mbox{tot}}<u_{2s}.
\]
Since $v$ changes sign when $u$ passes $E_{\mbox{tot}}/F_{\mbox{tot}}$, so the corresponding $v_{2t+1}$ and $v_{2s}$ have different signs, and either $(u_{2t+1},v_{2t+1}) \notin \mathcal S(\kappa, \mathcal C)$ or $(u_{2s},v_{2s}) \notin \mathcal S(\kappa, \mathcal C)$, contradiction.
\end{enumerate}
Therefore, $\Sigma(\kappa, \mathcal C)$ has at most $2n-1$ steady states.
\end{proof}

\subsection{Fine-tuned upper bounds}

In the previous section, we have seen that any $(u,v) \in \mathcal S(\kappa, \mathcal C), u \not= E_{\mbox{tot}}/F_{\mbox{tot}}$ must satisfy \eqref{eqn:vv}-\eqref{eqn:uu}, 
but not all solutions of \eqref{eqn:vv}-\eqref{eqn:uu} are elements in $\mathcal S(\kappa, \mathcal C)$.
Suppose that $(u,v)$ is a solution of \eqref{eqn:vv}-\eqref{eqn:uu},
it is in $\mathcal S(\kappa, \mathcal C)$ if and only if $u,v>0$. In some special cases, for example, when the enzyme is in excess, or the substrate is in excess, we could count the number of solutions of \eqref{eqn:vv}-\eqref{eqn:uu} which are not in $\mathcal S(\kappa, \mathcal C)$ to get a better upper bound. 

The following is a standard result on continuity of roots; see for instance Lemma A.4.1 in \cite{mct}:
\begin{lemma}
\label{lemma:Sontag}
Let $g(z)=z^n+a_1 z^{n-1}+ \cdots + a_n $ be a polynomial of degree $n$ and complex coefficients having distinct roots 
\[
\lambda_1, \dots, \lambda_q,
\]
with multiplicities 
\[
n_1+\cdots + n_q=n,
\]
respectively. Given any small enough $\delta>0$ there exists a $\varepsilon>0$ so that, if 
\[
h(z)=z^n+b_1 z^{n-1}+ \cdots + b_n, \ \ |a_i-b_i|<\varepsilon \mbox{ for } i=1,\dots, n,
\]
then $h$ has precisely $n_i$ roots in $B_\delta(\lambda_i)$ for each $i=1,\dots, q$.
\end{lemma}
\begin{theorem}
\label{thm:n+1}
For each $\gamma>0$ and $\kappa \in \mathbb R^{6n-6}_+$ such that $\varphi_1^\kappa(\gamma)\not=\varphi_2^\kappa(\gamma)$, and each $S_{\mbox{tot}}>0$, there exists $\varepsilon_2>0$ such that for all positive numbers $E_{\mbox{tot}}, F_{\mbox{tot}}$ satisfying $F_{\mbox{tot}}=E_{\mbox{tot}}/\gamma<\varepsilon_2 S_{\mbox{tot}}/\gamma$, the system $\Sigma(\kappa, \mathcal C)$ has at most $n+1$ positive steady states.
\end{theorem}
\begin{proof}
Let us define a function $\mathbb R_+ \times \mathbb C \longrightarrow \mathbb C$ as follows:
\[
\tilde Q^{\kappa, \gamma, S_{\mbox{tot}}}(\varepsilon,u)=Q^{\kappa, (\varepsilon S_{\mbox{tot}},  \varepsilon S_{\mbox{tot}}/\gamma, S_{\mbox{tot}})}(u),
\]
and a set $\mathcal B^{\kappa, \gamma, S_{\mbox{tot}}}(\varepsilon)$ consisting of the roots of $\tilde Q^{\kappa, \gamma, S_{\mbox{tot}}}(\varepsilon,u)$ which are not positive or the corresponding $v$'s determined by $u$'s as in \eqref{eqn:vv} are not positive, 
Since $\tilde Q^{\kappa, \gamma, S_{\mbox{tot}}}(\varepsilon,u)$ is a polynomial of degree 
$2n+1$, if we can show that there exists $\varepsilon_2>0$ such that for any $\varepsilon \in (0, \varepsilon_2)$, $\tilde Q^{\kappa, \gamma, S_{\mbox{tot}}}(\varepsilon,u)$ has at least 
$n$ roots counting multiplicities  that are in $\mathcal B^{\kappa, \gamma, S_{\mbox{tot}}}(\varepsilon)$, then we are done.

In order to apply Lemma \ref{lemma:Sontag}, we regard the function $\tilde Q^{\kappa, \gamma, S_{\mbox{tot}}}$ as defined on $\mathbb R \times \mathbb C$. At $\varepsilon=0$:
\begin{align*}
\tilde Q^{\kappa, \gamma, S_{\mbox{tot}}}(0,u)&=[\varphi_0^\kappa\varphi_2^\kappa(\gamma-u)^2+(\varphi_0^\kappa-S_{\mbox{tot}}\varphi_2^\kappa)(u\varphi_1^\kappa-\gamma \varphi_2^\kappa)(\gamma-u)-S_{\mbox{tot}}(u\varphi_1^\kappa-\gamma \varphi_2^\kappa)^2]/u\\
&=[\varphi_0^\kappa\varphi_2^\kappa(\gamma-u)^2+\varphi_0^\kappa(u\varphi_1^\kappa-\gamma \varphi_2^\kappa)(\gamma-u)-S_{\mbox{tot}}\varphi_2^\kappa(u\varphi_1^\kappa-\gamma \varphi_2^\kappa)(\gamma-u)-S_{\mbox{tot}}(u\varphi_1^\kappa-\gamma \varphi_2^\kappa)^2]/u\\
&=[\varphi_0^\kappa(\gamma-u)u(\varphi_1^\kappa-\varphi_2^\kappa)+S_{\mbox{tot}}u(u\varphi_1^\kappa-\gamma \varphi_2^\kappa)(\varphi_2^\kappa-\varphi_1^\kappa)]/u\\
&=(\varphi_2^\kappa-\varphi_1^\kappa)(u\varphi_0^\kappa+S_{\mbox{tot}}(u\varphi_1^\kappa-\gamma \varphi_2^\kappa)-\gamma \varphi_0^\kappa)\\
&=(\varphi_2^\kappa-\varphi_1^\kappa)\tilde F^{\kappa, \gamma, S_{\mbox{tot}}}(0,u)
\end{align*}
Let us denote the distinct roots of $\tilde Q^{\kappa, \gamma, S_{\mbox{tot}}}(0,u)/u$ as
\[
u_1, \dots, u_q,
\]
with multiplicities
\[
n_1+\cdots + n_q=2n+1,
\]
and the roots of $\varphi_1^\kappa-\varphi_2^\kappa$ as
\[
u_1,\dots, u_p, \ \ p \leq q,
\]
with multiplicities
\[
m_1+\cdots + m_p=n, \ \ n_i \geq m_i, \mbox{ for } i=1, \dots, p.
\]
For each $i=1, \dots, p$, if $u_i$ is real
and positive, then there are two cases ($u_i \not= \gamma$ as $\varphi_1^\kappa(\gamma) \not=\varphi_2^\kappa(\gamma)$):
\begin{enumerate}
\item $u_i>\gamma$. We have
\[
u_i\varphi_1^\kappa(u_i)-\gamma \varphi_2^\kappa(u_i)>\gamma(\varphi_1^\kappa(u_i)-\varphi_2^\kappa(u_i))=0.
\]
\item $u_i<\gamma$. We have
\[
u_i\varphi_1^\kappa(u_i)-\gamma \varphi_2^\kappa(u_i)<\gamma(\varphi_1^\kappa(u_i)-\varphi_2^\kappa(u_i))=0.
\]
\end{enumerate}

In both cases, $u_i\varphi_1^\kappa(u_i)-\gamma \varphi_2^\kappa(u_i)$ and $\gamma-u_i$ have opposite signs, i.e.
\[
(u_i\varphi_1^\kappa(u_i)-\gamma \varphi_2^\kappa(u_i))(\gamma-u_i)<0.
\]

Let us pick $\delta>0$ small enough such that the following conditions hold:
\begin{enumerate}
\item For all $i=1,\dots, p$, if $u_i$ is not real, then $B_\delta(u_i)$ has no intersection with the real axis.
\item For all $i=1,\dots, p$, if $u_i$ is real
and positive, the following inequality holds for any real $u \in B_\delta(u_i)$:
\begin{equation}
\label{eqn:negative_s_0}
(u\varphi_1^\kappa(u)-\gamma \varphi_2^\kappa(u))(\gamma-u)<0.
\end{equation}
\item For all $i=1,\dots, p$, if $u_i$ is real and negative, then $B_\delta(u_i)$ has no intersection with the imaginary axis.
\item $B_\delta(u_j)\bigcap B_\delta(u_k)=\emptyset$ for all $j \not=k=1, \dots, q$.
\end{enumerate}
By Lemma \ref{lemma:Sontag}, there exists $\varepsilon_3>0$ such that for all $\varepsilon \in (0, \varepsilon_3)$, the polynomial $\tilde Q^{\kappa, \gamma, S_{\mbox{tot}}}(\varepsilon,u)/u$ has exactly $n_j$ roots in each $B_\delta(u_j), j=1,\dots, q$, denoted by $u_j^k(\varepsilon), k=1,\dots, n_j$. 

We pick one such $\varepsilon$, and we claim that none of the roots in $B_\delta(u_i), i=1,\dots, p$ with the $v$ defined as in \eqref{eqn:vv} will be an element in $\mathcal S$. If so, we are done, since there are $\sum_1^p n_i \geq \sum_1^p m_i=n$ such roots,
of $\tilde Q^{\kappa, \gamma, S_{\mbox{tot}}}(\varepsilon,u)$ which are in $\mathcal B^{\kappa, \gamma, S_{\mbox{tot}}}(\varepsilon)$.

For each $i=1,\dots, p$, there are two cases:
\begin{enumerate}
\item $u_i$ is not real. Then condition 1 guarantees that $u_i^k(\varepsilon)$ is not real for each $k=1,\dots, n_i$, and thus is in $\mathcal B^{\kappa, \gamma, S_{\mbox{tot}}}(\varepsilon)$.

\item $u_i$ is real
and positive. Pick any root $u_i^k(\varepsilon)\in B_\delta(u_i), k=1,\dots, n_i$, the corresponding $v_i^k(\varepsilon)$ equals
\[
\frac{\gamma-u_i^k(\varepsilon)}{\big(u_i^k(\varepsilon)\varphi_1^\kappa(u_i^k(\varepsilon))-\gamma \varphi_2^\kappa(u_i^k(\varepsilon))\big)}<0
\]
followed from \eqref{eqn:negative_s_0}. So $(u_i^k(\varepsilon),v_i^k(\varepsilon))\notin \mathcal S(\kappa,\mathcal C)$, and $u_i^k(\varepsilon)\in \mathcal B^{\kappa, \gamma, S_{\mbox{tot}}}(\varepsilon)$.
\item $u_i$ is real and negative. By condition 1 and 3, $u_i^k(\varepsilon)$ is not positive for all $k=1,\dots, n_i$.
\end{enumerate}
\end{proof}

The next theorem considers the case when enzyme is in excess:
\begin{theorem}
\label{thm:1}
For each $\gamma>0, \kappa \in \mathbb R^{6n-6}_+$ such that $\varphi_1^\kappa(\gamma) \not=\varphi_2^\kappa(\gamma)$, and each $E_{\mbox{tot}}>0$, there exists $\varepsilon_3>0$ such that for all positive numbers $F_{\mbox{tot}}, S_{\mbox{tot}}$ satisfying $F_{\mbox{tot}}= E_{\mbox{tot}}/\gamma>S_{\mbox{tot}}/(\varepsilon_3\gamma)$, the system $\Sigma(\kappa,\mathcal C)$ has at most one positive steady state.
\end{theorem}
\begin{proof}
For each $\gamma>0, \kappa \in \mathbb R^{6n-6}_+$ such that $\varphi_1^\kappa(\gamma) \not=\varphi_2^\kappa(\gamma)$, and each $E_{\mbox{tot}}>0$, we define a function $\mathbb R_+ \times \mathbb C \longrightarrow \mathbb C$ as follows:
\[
\bar Q^{\kappa, \gamma, E_{\mbox{tot}}}(\varepsilon,u)=Q^{\kappa, (E_{\mbox{tot}}, E_{\mbox{tot}}/\gamma, \varepsilon E_{\mbox{tot}})}(u).
\]
Let us define the set $\mathcal C^{\kappa,\gamma,E_{\mbox{tot}}}(\varepsilon)$ as the set of roots of $\bar Q^{\kappa, \gamma, E_{\mbox{tot}}}(\varepsilon,u)$ which are not positive or the corresponding 
$v$'s determined by $u$'s as in \eqref{eqn:vv} are not positive. If we can show that there exists $\varepsilon_3>0$ such that for any $\varepsilon \in (0, \varepsilon_3)$ there is at most one positive root of $\bar Q^{\kappa, \gamma, E_{\mbox{tot}}}(\varepsilon,u)$ that is not in $\mathcal C^{\kappa,\gamma,E_{\mbox{tot}}}(\varepsilon)$, we are done.

In order to apply Lemma \ref{lemma:Sontag}, we now view the function $\bar Q^{\kappa, \gamma, E_{\mbox{tot}}}$ as defined on $\mathbb R \times \mathbb C$. At $\varepsilon=0$:
\begin{align*}
\bar Q^{\kappa, \gamma, E_{\mbox{tot}}}(0,u)&=\left(\gamma-u\right)\left(\left(\gamma-u\right)\varphi_0^\kappa\varphi_2^\kappa+\left(\varphi_0^\kappa+\frac{E_{\mbox{tot}}}{\gamma}u\varphi_1^\kappa+\frac{E_{\mbox{tot}}}{\gamma}\varphi_2^\kappa\right)\left(u\varphi_1^\kappa-\gamma \varphi_2^\kappa\right)\right)/u\\
&:=\left(\gamma-u\right)R^{\kappa, \gamma, E_{\mbox{tot}}}(u).
\end{align*}
Let us denote the distinct roots of $\bar Q^{\kappa, \gamma, E_{\mbox{tot}}}(0,u)/u$ 
as
\[
u_1(=\gamma), u_2, \dots, u_q,
\]
with multiplicities
\[
n_1+\cdots + n_q=2n+1,
\]
and $u_2,\dots, u_q$ are the roots of $R^{\kappa, \gamma, E_{\mbox{tot}}}(u)$ other than $\gamma$.

Since $\varphi_1^\kappa(\gamma) \not=\varphi_2^\kappa(\gamma)$, $R^{\kappa, \gamma, E_{\mbox{tot}}}(u)$ is not divisible by $u-\gamma$, and thus $n_1=1$.

For each $i=2, \dots, q$, we have 
\[
\left(\gamma-u_i\right)\varphi_0^\kappa(u_i)\varphi_2^\kappa(u_i)=-\left(\varphi_0^\kappa(u_i)+\frac{E_{\mbox{tot}}}{\gamma}u_i\varphi_1^\kappa(u_i)+\frac{E_{\mbox{tot}}}{\gamma}\varphi_2^\kappa(u_i)\right)\left(u_i\varphi_1^\kappa(u_i)-\gamma \varphi_2^\kappa(u_i)\right).
\]
If $u_i>0$, then $\varphi_0^\kappa(u_i)\varphi_2^\kappa(u_i)$ and $\varphi_0^\kappa(u_i)+\frac{E_{\mbox{tot}}}{\gamma}u_i\varphi_1^\kappa(u_i)+\frac{E_{\mbox{tot}}}{\gamma}\varphi_2^\kappa(u_i)$ are both positive. Since $u_i\varphi_1^\kappa(u_i)-\gamma \varphi_2^\kappa(u_i)$ and $\gamma-u_i$ are non zero, $u_i\varphi_1^\kappa(u_i)-\gamma \varphi_2^\kappa(u_i)$ and $\gamma-u_i$ must have opposite signs, that is
\[
(u_i\varphi_1^\kappa(u_i)-\gamma \varphi_2^\kappa(u_i))(\gamma-u_i)<0.
\]
Let us pick $\delta>0$ small enough such that the following conditions hold for all $i=2,\dots, q$:
\begin{enumerate}
\item If $u_i$ is not real, then $B_\delta(u_i)$ has no intersection with the real axis.
\item If $u_i$ is real and positive, then for any real $u \in B_\delta(u_i)$, the following inequality holds:
\begin{equation}
\label{eqn:negative_s_0_again}
(u\varphi_1^\kappa(u)-\gamma \varphi_2^\kappa(u))(\gamma-u)<0.
\end{equation}
\item If $u_i$ is real and negative, then $B_\delta(u_i)$ has no intersection with the imaginary axis.
\item $B_\delta(u_j)\bigcap B_\delta(u_k)=\emptyset$ for all $i\not=k=2,\dots, q$.
\end{enumerate}
By Lemma \ref{lemma:Sontag}, there exists $\varepsilon_3>0$ such that for all $\varepsilon \in (0, \varepsilon_3)$, the polynomial $\bar Q^{\kappa, \gamma,E_{\mbox{tot}}}(\varepsilon,u)$ has exactly $n_j$ roots in each $B_\delta(u_j), j=1,\dots, q$, denoted by $u_j^k(\varepsilon), k=1,\dots, n_j$. 

We pick one such $\varepsilon$, and if we can show that all of the roots in $B_\delta(u_i), i=2,\dots,q$ are in $\mathcal C^{\kappa,\gamma,E_{\mbox{tot}}}(\varepsilon)$, then we are done, since the only roots that may not be in $\mathcal C^{\kappa,\gamma,E_{\mbox{tot}}}(\varepsilon)$ are the roots in $B_\delta(u_1)$, and there is
one root in $B_\delta(u_1)$.

For each $i=2,\dots, p$, there are three cases:
\begin{enumerate}
\item $u_i$ is not real. Then condition 1 guarantees that $u_i^k(\varepsilon)$ is not real for all $k=1,\dots, n_i$.
\item $u_i$ is real and positive. Pick any root $u_i^k(\varepsilon), k=1,\dots, n_i$, the corresponding $v_i^k(\varepsilon)$ equals
\[
\frac{\gamma-u_i^k(\varepsilon)}{u_i^k(\varepsilon)\varphi_1^\kappa(u_i^k(\varepsilon))-\gamma \varphi_2^\kappa(u_i^k(\varepsilon))}<0.
\]
So, $u_i^k(\varepsilon)$ is in $\mathcal C^{\kappa,\gamma,E_{\mbox{tot}}}(\varepsilon)$.
\item $u_i$ is real and negative. By conditions 1 and 3, $u_i^k(\varepsilon)$ is not positive for all $k=1,\dots, n_i$.
\end{enumerate}
\end{proof}

\section{Conclusions and discussions}
\label{s:end}
Here we have set up a mathematical model for multisite phosphorylation-dephosphorylation cycles of size $n$, and studied the number of positive steady states based on this model. We reformulated the question of number of positive steady states to question of the number of positive roots of certain polynomials, through which we also applied perturbation techniques. Our theoretical results depend on the assumption of mass action kinetics and distributive sequential mechanism, which are customary in the study of multisite phosphorylation and dephosphorylation.

An upper bound of $2n-1$ steady states is obtained for arbitrary parameter combinations. Biologically, when the substrate concentration greatly exceeds that of the enzyme, there are at most $n+1$ ($n$) steady states if $n$ is even (odd). And this upper bound can be achieved under proper kinetic conditions, see Theorem \ref{thm:exist} for the construction. On the other extreme, when the enzyme is in excess, there is a unique steady state. 

As a special case of $n=2$, which can be applied to a single level of MAPK cascades. Our results guarantees that there are no more than three steady states, consistent with numerical simulations in \cite{Kholodenko}.

We notice that there is an apparent gap between the upper bound $2n-1$ and the upper bound of $n+1$ ($n$) if $n$ is even (odd) when the substrate is in excess. If we think the ratio $E_{\mbox{tot}}/F_{\mbox{tot}}$ as a parameter $\varepsilon$, then when $\varepsilon \ll 1$, there are at most $n+1$ ($n$) steady states when $n$ is even (odd), which coincides with the largest possible lower bound. When $\varepsilon \gg 1$, there is a unique steady state. If the number of steady states changes ``continuously'' with respect to $\varepsilon$, then we do not expect the number of steady states to exceed $n+1$ ($n$) if $n$ is even (odd). So a natural conjecture would be that the number of steady states never exceed $n+1$ under any conditions.

\section{Acknowledgment}

We thank Jeremy Gunawardena for very helpful discussions.
\section{Appendix}
{\em proof of Lemma \ref{lemma:derivative}}:
Recall that (dropping the $u$'s in $\varphi_i^\kappa, i=0,1,2$)
\[
\tilde F^{\kappa, \gamma, S_{\mbox{tot}}}(0,u)=u\varphi_0^\kappa+S_{\mbox{tot}}(u\varphi_1^\kappa-\gamma \varphi_2^\kappa)-\gamma \varphi_0^\kappa.
\]
So
\begin{align*}
\frac{\partial \tilde F^{\kappa, \gamma, S_{\mbox{tot}}}}{\partial u}(0,u)&=\varphi_0^\kappa+S_{\mbox{tot}}(u\varphi_1^\kappa-\gamma \varphi_2^\kappa)'-(\gamma-u)(\varphi_0^\kappa)'.
\end{align*}
Since $\tilde F^{\kappa, \gamma, S_{\mbox{tot}}}(0,\bar u)=0$,
\[
S_{\mbox{tot}}(\bar u \varphi_1^\kappa-\gamma \varphi_2^\kappa)=(\gamma-\bar u)\varphi_0^\kappa,
\]
that is,
\[
\gamma-\bar u=\frac{S_{\mbox{tot}}(\bar u \varphi_1^\kappa-\gamma \varphi_2^\kappa)}{\varphi_0^\kappa}.
\]
Therefore,
\begin{align*}
\frac{\partial \tilde F^{\kappa, \gamma, S_{\mbox{tot}}}}{\partial u}(0,\bar u) &=\varphi_0^\kappa+S_{\mbox{tot}}(u\varphi_1^\kappa-\gamma \varphi_2^\kappa)'-\frac{S_{\mbox{tot}}(\bar u \varphi_1^\kappa-\gamma \varphi_2^\kappa)}{\varphi_0^\kappa}(\varphi_0^\kappa)'\\
&=\varphi_0^\kappa+\frac{S_{\mbox{tot}}}{\varphi_0^\kappa}\left(\varphi_0^\kappa(u \varphi_1^\kappa-\gamma \varphi_2^\kappa)'-(\bar u \varphi_1^\kappa-\gamma \varphi_2^\kappa)(\varphi_0^\kappa)'\right) \\
&=\varphi_0^\kappa+\frac{S_{\mbox{tot}}}{\varphi_0^\kappa}((1+\lambda_0 \bar u+\lambda_0\lambda_1 \bar u^2+ \cdots +\lambda_0 \cdots \lambda_{n-1} \bar u^n)\times \\
&\left(\frac{1}{K_{M_0}}(1-\gamma \beta_0)+2\frac{\lambda_0}{K_{M_1}}(1-\gamma \beta_1)\bar u+\cdots +n\frac{\lambda_0\cdots\lambda_{n-2}}{K_{M_{n-1}}}(1-\gamma \beta_{n-1}) \bar u^{n-1}\right)\\
&-\left(\lambda_0+2\lambda_0\lambda_1 \bar u+ \cdots +n \lambda_0 \cdots \lambda_{n-1} \bar u^{n-1}\right)\times \\
&\left(\frac{1}{K_{M_0}}(1-\gamma \beta_0)\bar u +\frac{\lambda_0}{K_{M_1}}(1-\gamma \beta_1)\bar u^2+\cdots +\frac{\lambda_0\cdots\lambda_{n-2}}{K_{M_{n-1}}}(1-\gamma \beta_{n-1}) \bar u^{n}\right))\\
&=\varphi_0^\kappa+\frac{S_{\mbox{tot}}}{\varphi_0^\kappa}\sum_{i=0}^n \lambda_0\cdots\lambda_{i-1} \bar u^i\left(\sum_{j=0}^{n-1}(j+1-i)\frac{\lambda_0 \cdots \lambda_{j-1}}{K_{M_{j}}}(1-\gamma \beta_{j})\bar u^j \right)\\
&=\frac{1}{\varphi_0^\kappa}\sum_{i=0}^n \lambda_0\cdots\lambda_{i-1} \bar u^i\sum_{j=0}^n \lambda_0\cdots\lambda_{j-1} \bar u^j\\
&+S_{\mbox{tot}}\sum_{i=0}^n \lambda_0\cdots\lambda_{i-1} \bar u^i\left(\sum_{j=0}^{n-1}(j+1-i)\frac{\lambda_0 \cdots \lambda_{j-1}}{K_{M_{j}}}(1-\gamma \beta_{j})\bar u^j )\right)\\
&=\frac{1}{\varphi_0^\kappa}\sum_{i=0}^n \lambda_0\cdots\lambda_{i-1} \bar u^i\left(\lambda_0\cdots\lambda_{n-1} \bar u^{n}+\sum_{j=0}^{n-1} \lambda_0\cdots\lambda_{j-1} \bar u^j\left(1+S_{\mbox{tot}}(j+1-i)\frac{1-\gamma \beta_{j}}{K_{M_{j}}}\right)\right),
\end{align*}
where the product $\lambda_0\cdots \lambda_{-1}$ is defined to be $1$ for the convenience of notation.

Because of \eqref{eqn:condition},
\[
S_{\mbox{tot}}\left|(j+1-i)\frac{1-\gamma \beta_{j}}{K_{M_{j}}}\right|\leq 1,
\]
so we have $\frac{\partial \tilde F^{\kappa, \gamma, S_{\mbox{tot}}}}{\partial u}(0,\bar u) >0$.
{\hspace*{\fill}$\halmos$\medskip}
\newpage


\begin{thebibliography}{10}

\bibitem{Samoilov}
M.~Samoilov, S.~Plyasunov, and A.P. Arkin.
\newblock Stochastic amplification and signaling in enzymatic futile cycles
  through noise-induced bistability with oscillations.
\newblock {\em Proc Natl Acad Sci USA}, 102:2310--2315, 2005.

\bibitem{Donovan}
S.~Donovan, K.M. Shannon, and G.~Bollag.
\newblock {GTPase} activating proteins: critical regulators of intracellular
  signaling.
\newblock {\em Biochim. Biophys Acta}, 1602:23--45, 2002.

\bibitem{groisman}
J.J. Bijlsma and E.A. Groisman.
\newblock Making informed decisions: regulatory interactions between
  two-component systems.
\newblock {\em Trends Microbiol}, 11:359--366, 2003.

\bibitem{grossman}
A.D. Grossman.
\newblock Genetic networks controlling the initiation of sporulation and the
  development of genetic competence in bacillus subtilis.
\newblock {\em Annu Rev Genet.}, 29:477--508, 1995.

\bibitem{chen}
H.~Chen, B.W. Bernstein, and J.R. Bamburg.
\newblock Regulating actin filament dynamics in vivo.
\newblock {\em Trends Biochem. Sci.}, 25:19--23, 2000.

\bibitem{karp}
G.~Karp.
\newblock {\em Cell and Molecular Biology}.
\newblock Wiley, 2002.

\bibitem{stryer}
L.~Stryer.
\newblock {\em Biochemistry}.
\newblock Freeman, 1995.

\bibitem{sulis}
M.L. Sulis and R.~Parsons.
\newblock {PTEN}: from pathology to biology.
\newblock {\em Trends Cell Biol.}, 13:478--483, 2003.

\bibitem{lew}
D.J. Lew and D.J. Burke.
\newblock The spindle assembly and spindle position checkpoints.
\newblock {\em Annu Rev Genet.}, 37:251--282, 2003.

\bibitem{lauffenburger}
A.R. Asthagiri and D.A. Lauffenburger.
\newblock A computational study of feedback effects on signal dynamics in a
  mitogen-activated protein kinase (MAPK) pathway model.
\newblock {\em Biotechnol. Prog.}, 17:227--239, 2001.

\bibitem{Chang}
L.~Chang and M.~Karin.
\newblock Mammalian {MAP} kinase signaling cascades.
\newblock {\em Nature}, 410:37--40, 2001.

\bibitem{ferrell}
C-Y.F. Huang and J.E.~Ferrell Jr.
\newblock Ultrasensitivity in the mitogen-activated protein kinase cascade.
\newblock {\em Proc. Natl. Acad. Sci. USA}, 93:10078--10083, 1996.

\bibitem{widman}
C.~Widmann, G.~Spencer, M.B. Jarpe, and G.L. Johnson.
\newblock Mitogen-activated protein kinase: Conservation of a three-kinase
  module from yeast to human.
\newblock {\em Physiol. Rev.}, 79:143--180, 1999.

\bibitem{Burack}
W.R. Burack and T.W. Sturgill.
\newblock The activating dual phosphorylation of {MAPK} by {MEK} is
  nonprocessive.
\newblock {\em Biochemistry}, 36:5929--5933, 1997.

\bibitem{Ferrell_Bhatt}
J.E. Ferrell and R.R. Bhatt.
\newblock Mechanistic studies of the dual phosphorylation of mitogen-activated
  protein kinase.
\newblock {\em J. Biol. Chem.}, 272:19008--19016, 1997.

\bibitem{Zhao}
Y.~Zhao and Z.Y. Zhang.
\newblock The mechanism of dephosphorylation of extracellular signal-regulated
  kinase 2 by mitogen-activated protein kinase phosphatase 3.
\newblock {\em J. Biol. Chem.}, 276:32382--32391, 2001.

\bibitem{Kholodenko}
N.I. Markevich, J.B. Hoek, and B.N. Kholodenko.
\newblock Signaling switches and bistability arising from multisite
  phosphorylation in protein kinase cascades.
\newblock {\em J. Cell Biol.}, 164:353--359, 2004.

\bibitem{jeremy}
J.~Gunawardena.
\newblock Multisite protein phosphorylation makes a good threshold but can be a
  poor switch.
\newblock {\em Proc. Natl. Acad. Sci.}, 102:14617--14622, 2005.

\bibitem{conradi}
C.~Conradi, J.~Saez-Rodriguez, E.-D. Gilles, and J.~Raisch.
\newblock Using chemical reaction network theory to discard a kinetic mechanism
  hypothesis.
\newblock In {\em Proc. FOSBE 2005 (Foundations of Systems Biology in
  Engineering), Santa Barbara, Aug. 2005}, pages 325--328. 2005.

\bibitem{Gardner}
T.S. Gardner, C.R. Cantor, and J.J. Collins.
\newblock Construction of a genetic toggle switch in {E}scherichia coli.
\newblock {\em Nature}, 403:339--342, 2000.

\bibitem{pnasangeliferrellsontag04}
D.~Angeli, J.~E. Ferrell, and E.D. Sontag.
\newblock Detection of multistability, bifurcations, and hysteresis in a large
  class of biological positive-feedback systems.
\newblock {\em Proc Natl Acad Sci USA}, 101(7):1822--1827, 2004.

\bibitem{Selkov}
E.E. Sel'kov.
\newblock Stabilization of energy charge, generation of oscillation and
  multiple steady states in energy metabolism as a result of purely
  stoichiometric regulation.
\newblock {\em Eur. J. Biochem}, 59(1):151--157, 1975.

\bibitem{Sha}
W.~Sha, J.~Moore, K.~Chen, A.D. Lassaletta, C.S. Yi, J.J. Tyson, and J.C.
  Sible.
\newblock Hysteresis drives cell-cycle transitions in {Xenopus} laevis egg
  extracts.
\newblock {\em Proc. Natl. Acad. Sci.}, 100:975--980, 2003.

\bibitem{Kholodenko_steady_state}
F.~Ortega, J.~Garc\'es, F.~Mas, B.N. Kholodenko, and M.~Cascante.
\newblock Bistability from double phosphorylation in signal transduction:
  Kinetic and structural requirements.
\newblock {\em FEBS J}, 273:3915--3926, 2006.

\bibitem{IEEEsysbio_WS}
L.~Wang and E.D. Sontag.
\newblock Singularly perturbed monotone systems and an application to double
  phosphorylation cycles.
\newblock (Submitted to IEEE Transactions Autom. Control, Special Issue on
  Systems Biology, January 2007, Preprint version in arXiv math.OC/0701575, 20
  Jan 2007), 2007.

\bibitem{06posta_wang}
L.~Wang and E.D. Sontag.
\newblock Almost global convergence in singular perturbations of strongly
  monotone systems.
\newblock In {\em Positive Systems}, pages 415--422. Springer-Verlag,
  Berlin/Heidelberg, 2006.
\newblock (Lecture Notes in Control and Information Sciences Volume 341,
  Proceedings of the second Multidisciplinary International Symposium on
  Positive Systems: Theory and Applications (POSTA 06) Grenoble, France).

\bibitem{persistencePetri}
D.~Angeli, P.~de~Leenheer, and E.D. Sontag.
\newblock A {P}etri net approach to the study of persistence in chemical
  reaction networks.
\newblock (Submitted to Mathematical Biosciences, also arXiv q-bio.MN/068019v2,
  10 Aug 2006), 2007.

\bibitem{translation-invariance}
D.~Angeli and E.D. Sontag.
\newblock Translation-invariant monotone systems, and a global convergence
  result for enzymatic futile cycles.
\newblock {\em Nonlinear Analysis Series B: Real World Applications}, to
  appear, 2007.

\bibitem{jeremy2}
M~Thompson and J.~Gunawardena.
\newblock Multi-bit information storage by multisite phosphorylation.
\newblock Submitted, 2007.

\bibitem{mct}
E.D. Sontag.
\newblock {\em Mathematical {C}ontrol {T}heory. {D}eterministic
  {F}inite-{D}imensional {S}ystems}, volume~6 of {\em Texts in Applied
  Mathematics}.
\newblock Springer-Verlag, New York, second edition, 1998.

\bibitem{Feinberg}
M.~Feinberg.
\newblock Chemical reaction network structure and the stability of complex isothermal reactors: II. Multiple steady states for networks of deficiency one.
\newblock {\em Chem. Eng. Sci.}, 43,1--25, 1988.

\bibitem{Ellison_Feinberg}
P.~Ellison, M.~Feinberg.
\newblock How catalytic mechanisms reveal themselves in multiple steady-state data: I. Basic principles.
\newblock {\em J. Symbolic Comput.}, 33, 275--305, 2002.

\bibitem{Furdui}
C.M.~Furdui, E.D.~Lew, J.~Schlessinger, K.S. Anderson.
\newblock Autophosphorylation of FGFR1 kinase is mediated by a sequential and precisely ordered reaction.
\newblock {\em Molecular Cell}, 21, 711--717, 2006.

\end{thebibliography}

\end{document}